\documentclass[twocolumn,showpacs,preprintnumbers,amsmath,amssymb]{revtex4}
\usepackage{graphicx}
\usepackage{dcolumn}
\usepackage{bm}
\def\bea {\begin{eqnarray}}
\def\eea {\end{eqnarray}}
\def\ra {\rightarrow}
\def\be {\begin{equation}}
\def\ee {\end{equation}}

\begin{document}
\title{Electromagnetic signals from Au+Au collisions at RHIC energy 
$\sqrt{s_{NN}}$=200 GeV and Pb+Pb collisions at LHC energy, 
$\sqrt{s_{NN}}$=2.76 TeV.}

\author{Jajati K. Nayak and Bikash Sinha }

\medskip

\affiliation{Variable Energy Cyclotron Centre, 1/AF, Bidhan Nagar, 
Kolkata - 700064}.

\date{\today}

\begin{abstract}

We analyse the recently available experimental data on direct photon 
productions from Au+Au collisions at $\sqrt{s_{NN}}$=200 GeV RHIC 
and from Pb+Pb collisions at $\sqrt{s_{NN}}$=2.76 TeV LHC energies. 
The transverse momentum ($p_T$) distributions have been evaluated with 
the assumption of an initial quark gluon plasma phase at  temperatures 
$T_i$=404 and 546 MeV with initial thermalisation times $\tau_i$=0.2 and 
0.1 fm/c respectively for RHIC and LHC energies. The theoretical 
evalutions agree reasonably well with the experimental observations.
The thermal window for the LHC energy is justified through the ratio 
of the $p_T$ spectra of thermal photons to dileptons.
\end{abstract}

\pacs{25.75.-q,25.75.Dw,24.85.+p}
\maketitle

It is by now conventional wisdom, that collisions between two nuclei at 
RHIC as well as LHC energies will lead to the formation of quark gluon 
plasma (QGP). Although the exact nature of QGP-hadron phase transition 
is still plagued by uncertainties, but there are several evidences 
that QGP is formed~\cite{brahmswhitepaper,phoboswhitepaper,starwhitepaper,
phenixwhitepaper} at RHIC energies. 
It has long been recognised that promising signals of QGP are 
photons and dileptons~\cite{larry,GK,weldon90,jpr,annals}. The very 
nature of electromagnetic interactions ensures that the thermometric 
signals ($\gamma, \mu^+\mu^-, e^+e^-$) escape the QGP medium without 
any significant interaction, thus retaining the pristine information of 
QGP immediately after its formation. It has been pointed out by 
one of the authors and his collaborators, consistently for a long time 
~\cite{jpr,annals,assb,rsas,ashns,JA07,pm2010}, that the photons 
emanating from thermal medium show up in the $p_T$ window of 
$1.5 \le p_T (GeV) \le 3.5$. It has also been pointed out ~\cite{bsinha,ds} 
during the early days of QGP physics and now, that the ratio 
$\gamma/\mu^+\mu^-$ eliminates some of the uncertainties associated with 
the input parameters used in the model. 
\par
The concept of ratio has been revisited again in 
ref~\cite{JKN, JKN3} where the sensitivities of the of the initial 
parameters have been studied extensively. The formation of a thermal medium 
is manifested by the flattering of the ratio beyond certain $p_T$. 
Now that the direct photon data has come out of LHC we tend to analyse the 
data and test the above prediction at $\sqrt{s_{NN}}$=2.76 TeV. 
\par
It is quite remarkable that the original theoretical prediction
~\cite{bsinha,ds,JKN,JKN3} turns out to be valid also at LHC energies with 
added advantage that the very high energy of LHC turns the invariant mass $M$, 
of dileptons less relevant compare to (say) at RHIC energies. 
\par
It should also be noted that there is already a tantalising hint of QGP 
formation 
even at SPS energy~\cite{na60,ahns}. In the following work we present our 
findings of this interesting phenomenon. Thus over a wide range of energies 
going through orders of magnitude, it is quite exciting to note that $p_T$ 
window for thermal photons and dileptons remains the same as predicted 
earlier, only the range of window increases almost upto 4 GeV or more at 
LHC energy. 
\par
In this present work we have evaluated the direct photon productions 
from (i) Au+Au collisions at $\sqrt{s_{NN}}$=200 GeV for 0-5\% 
centrality and (ii) Pb+Pb collisions at $\sqrt{s_{NN}}$=2.76 TeV 
for 0-40\% centrality and finally compared with the recently 
available data ~\cite{phenix12}for RHIC and ~\cite{alicephoton} for 
LHC energies. We analyse the data using our earlier 
approach, which has been used consistently to explain the SPS 
\cite{pm2010} and RHIC ~\cite{JA07} data for different centralities. 
The invariant thermal yield of photons is evaluated from the production rate 
as in ref~\cite{JKN,JKN2} and using the (2+1) dimensional relativistic 
hydrodynamics. The photon contributions from prompt productions 
have been evaluated using the parton distribution function as parametrised in
{\it Cteq6M}. Different thermal sources of photon productions are  
considered as in ref~\cite{JKN,JKN2} and sources of prompt 
productions as in ref~\cite{assb,gordon}. The authors in 
~\cite{turbzakha} also mention the contribution of jet-plasma interaction 
to the high $p_T$ photons. Here the data seems to be explained nicely 
without considering the above phenomenon.
The dilepton productions are evaluated as in ~\cite{ahns} using 
the equation of state mentioned in sec-II. Then we evaluate ratio of 
the transverse momentum spectra of thermal photons to dileptons to seek out 
the thermal window at $\sqrt{s_{NN}}$=2.76 TeV.

\par 
In section II the prompt and thermal photon productions have been dscussed. 
In the following section the expansion dynamics of the produced system, 
along with the initial conditions, is described. Finally, the results have 
been discussed and compared with the available data. The summary is presented 
last.
\par
On the basis of sources of production,photons emanating from the relativistic 
heavy ion collision, can be broadly categorized as follows; (i) prompt photons 
resulting from the interactions of the partons of the colliding nuclei, 
(ii) pre-equilibrium photons, emitted before the medium gets thermalised, 
(iii) thermal photons originating  from the interaction of thermal partons 
as well as thermal hadrons, (iv) photons produced the from the passage of 
jets through plasma and (v) the photons from the decay of long lived hadrons 
($\pi^0 \rightarrow \gamma \gamma$, $\eta \rightarrow \gamma \gamma$, 
{\it etc}). The aim of this work is to analyse the photon data from 
RHIC and LHC energies, where photons from the hadronic decays, have already 
been excluded. Out of the remaining categories, the pre-equilibrium 
contributions is negligible since thermalisation times($\tau_i$) for RHIC 
and LHC are very small. The photons due to the passage of jets through the 
medium is not considered here. In this work, we consider the prompt and 
thermal photons-the two main sources of direct photons, in terms of their 
transverse momentum ($p_T$) distribution. 
\par
The prompt photons originate from the initial hard scatterings of the 
partons, primarily because of the compton processes 
($q (\bar q)+g \rightarrow q (\bar q)+\gamma$), annihilation process 
($q +\bar q \rightarrow g +\gamma$) and quark fragmentation 
($q (\bar q) \rightarrow q (\bar q) +\gamma$) following the scattering 
of partons of the nucleus in the colliding nuclei. In a complete and 
consistent NLOpQCD [$O(\alpha\alpha_s^2)$] calculation, it is important 
to account for the prompt photons arising from various $2\rightarrow 3$ 
processes~\cite{gordon1,aurenchea}. This calculation is comparable with the 
leading order calculation~\cite{assb} with a higher $K$ factor. We use this 
calculation and scale it up by the number of binarry collisions for 
Au+Au and Pb+Pb collisions at $\sqrt{s_{NN}}$=200 GeV (RHIC) and 
2.76 TeV(LHC) energies respectively. The intrinsic $k_T$ smearing 
is ignored here.
\par
The $p_T$ distribution of thermal photons is expressed in terms of 
invariant yield as follows; 
\begin{equation}
\frac{d^2N_{\gamma}}{d^2p_Tdy}=\sum_{i=phases}
\int_i{\left(\frac{d^2R_{\gamma}}{d^2p_Tdy}\right)_id^4x}
\label{inyield}
\end{equation}
$N_{\gamma}$ is the number of photons produced. $y$ is the rapidity.
where $i$ represents QGP and Hadronic phases. 
$\left(\frac{d^2R_{\gamma}}{d^2p_Tdy}\right)_i$ is the static rate of 
photon production at a temperature $T$ from phase $i$. $d^4x$ is 
the four volume element and its evolution is taken care by relativistic 
ideal hydrodynamics. The photon emission rate is given as follows;
\begin{equation}
\frac{d^2R}{d^2p_Tdy}=\frac{g^{\mu\nu}}{(2\pi)^3}\mathrm{Im\Pi_{\mu\nu}}f_{BE}
\label{real}
\end{equation}
(see~~\cite{larry,GK,weldon90,jpr,annals} for details). 
$\mathrm{Im\Pi_{\mu\nu}}$ is the imaginary part of the retarded photon self 
energy. $f_{BE}$ is the thermal distribution function. 
($g^{\mu\nu}$=-$\sum_\mathrm{polarization}\epsilon^\mu\epsilon^\nu$).
The static emission rate of photons for QGP and hadronic phases are 
calculated from various partonic and hadronic processes.
\par
The photon emission rate from QGP due to compton 
($q(\bar{q})g\rightarrow q(\bar{q})\gamma$) 
and annihilation ($q\bar{q}\rightarrow g\gamma$) processes 
were evaluated~\cite{kapusta,baier} by using hard thermal loop (HTL) 
approximation~\cite{pisarski}. Further it was found~\cite{auranche1}
that photon productions from the reactions,
$gq\rightarrow gq\gamma$, $qq\rightarrow qq\gamma$,
$qq\bar{q}\rightarrow q\gamma$ 
and $gq\bar{q}\rightarrow g\gamma$ contribute in the same order
as annihilation and compton processes do.
The suppression due to multiple scattering during the
emission process which was not mentioned in ~\cite{kapusta,baier,auranche1} 
was later discussed in Ref.~\cite{auranche2}. The complete calculation of 
photon emission rate from QGP to order ~O($\alpha\alpha_s$) has been 
completed by resuming ladder diagrams in the effective theory~\cite{arnold}, 
which has been used in the present work. The emission rates for various 
processes are available in parametrised form in Ref.~\cite{renk}. 
Here we use the temperature dependence of the strong coupling constant 
from Ref.~\cite{zantow}.
\begin{figure}
\begin{center}
\includegraphics[scale=0.43]{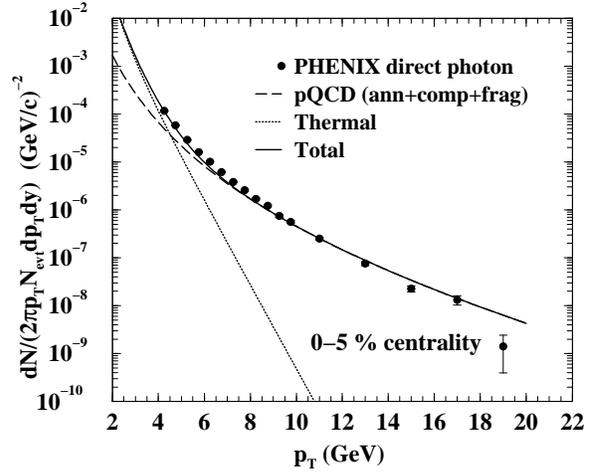}
\caption{The $p_T$ spectra of direct photons from Au+Au collisions 
at $\sqrt{s_{NN}}$=200 GeV RHIC energy from different sources. 
} 
\label{fig1}
\end{center}
\end{figure} 

An exhaustive set of hadronic interactions have been considered to 
evaluate the photon emission rate from a hadronic phase. The relevant 
processes are basically;  
(i) $\pi\,\pi\,\ra\,\rho\,\gamma$, (ii) $\pi\,\rho\,
\ra\,\pi\gamma$ (with $\pi$, $\rho$, $\omega$, $\phi$ and $a_1$ in the
intermediate state~\cite{we3}), (iii)$\pi\,\pi\,\ra\,\eta\,\gamma$ and 
(iv) $\pi\,\eta\,\ra\,\pi\,\gamma$. Also the radiative decay of higher 
resonance states ~\cite{we1,we2,we3,we4} such as, 
$\rho\,\ra\,\pi\,\pi\,\gamma$ and $\omega\,\ra\,\pi\,\gamma$ produces 
photons. The corresponding vertices can be obtained
from various phenomenological Lagrangians as described in detail 
in Ref.~\cite{we1,we2,we3,we4}. The contributions from the 
reactions involving strange mesons like 
$\pi K^* \ra K\gamma$, $\pi K\ra K^*\gamma$, $\rho K\ra K\gamma$ and 
$KK^*\ra K\gamma$, have  been pointed out in ~\cite{turbide}
Contributions from other decays, such as 
$K^{\ast}(892)\,\ra\, K\,\gamma$, $\phi\,\ra\,\eta\,\gamma$, 
$b_1(1235)\,\ra\,\pi\,\gamma$, $a_2(1320)\,\ra\,\pi\,\gamma$ 
and $K_1(1270)\,\ra\,\pi\,\gamma$ have been found to be 
small~\cite{haglin} for $p_T>1$ GeV. We consider all the 
processes for our calculation. All the isospin combinations for the 
above reactions and decays have properly been taken into account.
The effects of hadronic form factors(dipole) ~\cite{turbide} have 
been incorporated.
\par
When two energetic heavy nuclei collide with each other, a large 
amount of energy and thus entropy is channeled into a small volume. 
The produced system achieve thermalisation in a short time because 
of the secondary interactions and 
the matter then expands due to the high internal pressure.  
The space-time evolution of the matter has been studied using ideal 
relativistic hydrodynamics ~\cite{von} with longitudinal boost 
invariance ~\cite{bjorken} and cylindrical symmetry. 
In the present analysis of Pb+Pb collisions we consider the initial energy 
density ($\epsilon(\tau_i,r)$) and radial velocity
($v(\tau_i,r)$) profiles similar to our earlier studies
~\cite{ahns,JKN,JKN2,pm2010}. Transition temperature which is an 
input parameter to the hydrodynamic calculation is taken as $T_c$=163 MeV 
according the recent works of lattice QCD computation by the HotQCD 
Collaboration ~\cite{hotqcd} and by the Bielefield Group~\cite{bielefield}. 
We also consider $T_c$=170 MeV as predicted by the lattice 
computation from other group~\cite{latticetifr} to compare the results. 
The initial temperature $T_i$ and initial thermalisation time $\tau_i$ 
are constrained from the following equation~\cite{hwa};
\begin{equation}
T_i^3 (b) \equiv \frac{2 \pi^4}{45 \zeta (3)}
         \frac{1}{\pi R^2 \tau _i}
         \frac{90}{4\pi^2g_{eff}}\frac{dN}{dy}(b),
\label{intime}
\end{equation}

where $\zeta(3)$ denotes the Riemann zeta function, $R$ is the  
radius [$\sim$ $1.1(N_{part}/2)^{1/3}$, $N_{part}$ is the number of
participant nucleons ] of the colliding system , ${\tau}_i$ is the initial
thermalisation time and $g_{eff}$ is the statistical degeneracy taken to 
be 37, considering a 2-flavor QGP to be produced initially. $dN/dy(b)$ is 
the hadron (mostly pion) multiplicity for a given impact parameter $b$, 
which is obtained 
from the Glauber model for 0-40\% collision centrality of Pb+Pb collisions 
at $\sqrt{s_{NN}}$=2.76 TeV. The initial temperature obtained from the 
above equation Eq.\ref{intime} for $dN/dy(b)$=1275 and $\tau_i$=0.1 fm 
is 546 MeV. 
\par
The heavy ion collision experiment at LHC energy maps the region 
, where the value of baryonic chemical potential, $\mu_B$ is small and 
the temperature is rather high. Guided by the lattice 
computation, which 
predicts a crossover from hadronic to partonic phase for $\mu_B$=0, 
we construct an equation of state (EOS) for the evolution dynamics which 
gives crossover like transition. We consider the entropy ($s_q$) of the 
QGP phase (for $T>T_c$) using BAG model EOS and the entropy ($s_h$) of the 
hadron gas (considering non-interacting hadrons and their resonances up 
to mass $\sim$ 2.5 GeV for $T<T_c$). The entropy during the transition region 
is parametrised using a tan-hyperbolic function~\cite{hatsuda} as follows.  
\begin{eqnarray}
s(T)=s_q(T) f_q(T)+[1-f_q(T)]s_h(T) \nonumber\\
\mbox{    and     } f_q(T)=\frac{1}{2}(1+tanh\frac{T-T_c}{\Gamma})
\label{entro}
\end{eqnarray}
Here $\Gamma$ is the width parameter and assumes a finite value for 
crosssover transition. This value can be tuned to zero to the 
first order transition. Here the width parameter is taken to be 
 $\Gamma$=20 MeV.
\begin{figure}
\begin{center}
\includegraphics[scale=0.43]{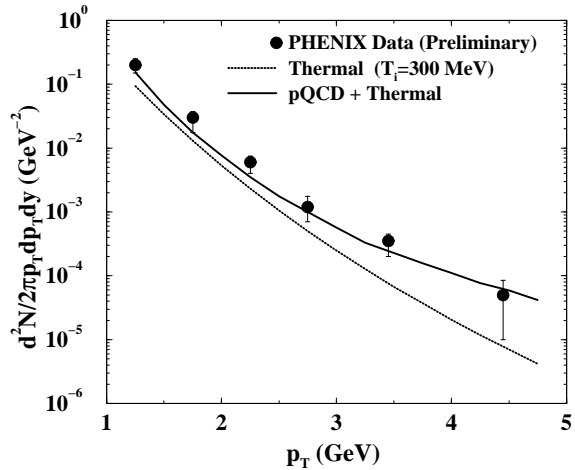}
\caption{ Direct photons measured by PHENIX collaboration 
for 0-20\% centrality at $\sqrt{s_{NN}}$=200 GeV~\cite{JA07}.  }
\label{fig2}
\end{center}
\end{figure} 
EOS which is an input to the hydrodynamic calculation has been constructed 
for a transition temperature $T_c$=163 MeV and width parameter 
$\Gamma$=20 MeV. The effective degeneracy $\mathrm g(T)$ of the 
system which is obtained from the entropy $s(T)$ (Eq.~\ref{entro}).
\par
The $p_T$ spectra of direct photons have been evaluated for Au+Au collisions 
at $\sqrt{s_{NN}}$=200 GeV RHIC energy. For 0-5\% of the centrality of the 
collision, the total pion multiplicity is taken as $dN/dy$=1100. To evaluate 
the thermal contribution, we consider an initial temperature $T_i$=404 MeV, 
which is obtained from the Eq.~\ref{intime} for an initial thermalisation 
time $\tau_i$=0.2 fm. An equal time freeze out scenario for all the hadrons 
is assumed and the value of freeze out temperature $T_f$ is taken to be 
120 MeV, which is contrasted from the pion and kaon spectra~\cite{JA07}. 
The prompt photon contributions have been evaluated from the annihilation, 
compton and quark-fragmaneation processes using pQCD calculations~\cite{assb}. 
The calculation is done for $N_{coll}$=910 (obtain from Glauber model 
calculation for 0-5\% centrality) and $\sigma_{in}$=42 mb. We use the 
parametrised parton distribution function from {\it Cteq6M }. To account 
for for the next-to-next leading order contribution, the value of 
$K_{\gamma}$ and $K_{brem}$ is taken to be 1.4 and 2.2 respectively for 
RHIC energy.  
The results are shown in Fig.~\ref{fig1}. The solid cirlces represent the 
recently published direct photon data from Au+Au collisions (0-5\% centrality) 
by PHENIX collaboration~\cite{phenix12}. The long-dashed and dotted lines 
represents the pQCD and thermal contributions respectively. The solid is for 
the sum total of both. $T_c$ is taken to be 163 MeV.
\begin{figure}
\begin{center}
\includegraphics[scale=0.43]{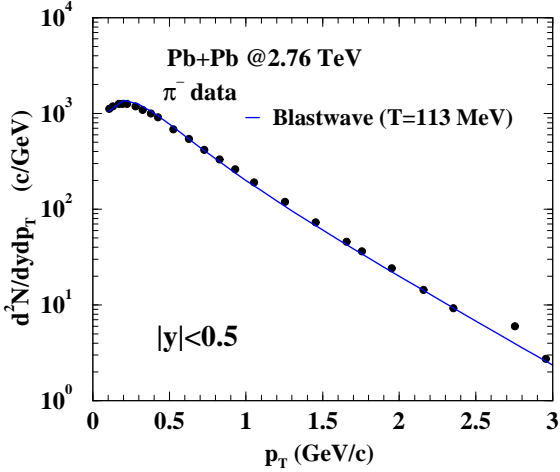}
\caption{The pion spectra from Pb+Pb collisions 
at $\sqrt{s_{NN}}$=2.76 TeV LHC energy fitted with blast wave with 
temperature $T_f$=113 MeV and $\beta$=0.9 }
\label{fig3}
\end{center}
\end{figure} 
\par
The Fig.~\ref{fig2} displays the direct photon productions for 0-20\% 
centrality of Au+Au collisions at RHIC energy~\cite{phenix06} as per our 
earlier evaluation~\cite{JA07}. 
The dotted line represents the thermal contribution for lattice EOS 
~\cite{karsch02} with $T_i$=300 MeV and $\tau_i$=0.5 fm. 
The prompt photons are evaluated by scaling the NLO calculation of Gordon and 
Vogelssang's ~\cite{gordon} prediction for p-p collisions with the number of 
collisions for Au+Au. The solid line represents the total photons.

\begin{figure}
\begin{center}
\includegraphics[scale=0.43]{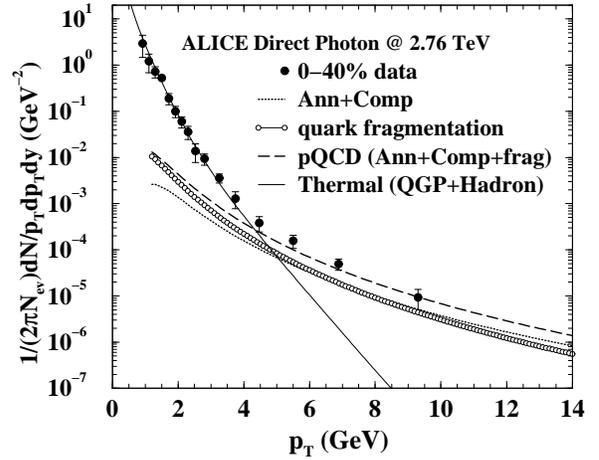}
\caption{The $p_T$ spectra of direct photons from Pb+Pb collisions 
at $\sqrt{s_{NN}}$=2.76 TeV LHC energy from different sources. 
} 
\label{fig4}
\end{center}
\end{figure}

While evaluating the $p_T$ distributions for direct photons at LHC energy,  
same time freezeout scenario for all the hadrons is assumed with 
$T_f$=113 MeV, which is constrained by the pion data. 
Fig.~\ref{fig3} displays the blast wave explanation of the pion data from Pb+Pb 
collisions at $\sqrt{s_{NN}}$=2.76 TeV (For data see ~\cite{alicepion} ) 
with $T$=113 MeV. \\
 In Fig.~\ref{fig4} the direct photon data from 
Pb+Pb collisions for 0-40\% centrality at $\sqrt{s_{NN}}$=2.76 TeV measured 
by ALICE collaboration (solid circles)~\cite{alicephoton} is shown. 
The dotted line represents the pQCD calculation of annihilation and 
compton processes (prompt) using {\it Cteq6m} for $N_{coll}$=853(for 0-40\%). 
The solid line with open circles show the contributions from 
quark-fragmentation to the prompt photons. 
The long-dashed line is the sum over all the hard processes. 
$K_{\gamma}$=3.2 for annihilation and compton processes 
and $K_{brem}$=2.8 for fragmentation process are considered to take the 
NNLO contributions into account.
\begin{figure}
\begin{center}
\includegraphics[scale=0.43]{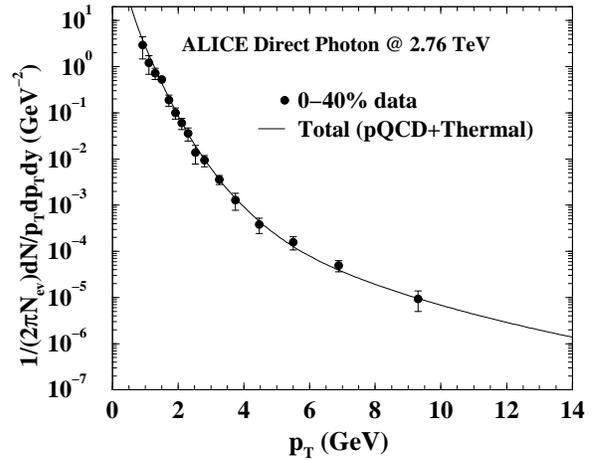}
\caption{The $p_T$ spectra of direct photons from Pb+Pb collisions 
at $\sqrt{s_{NN}}$=2.76 TeV LHC energy. The solid cirlces are the data 
points measured by ALICE collaborations~\cite{phenix12}. 
The solid line represents the theoretical evaluation. 
}
\label{fig5}
\end{center}
\end{figure} 
Considering $T_i$ to be 546 MeV, which is obtained from Eq.~\ref{intime} 
for 0-40\% centrality and $\tau_i$=0.1 fm, $T_f$=113 MeV and $T_c$=163 MeV, 
the thermal contributions are evaluated from QGP and hadron gas. This is 
displayed in Fig.~\ref{fig4} (solidline).
As depicted from Fig.~\ref{fig4}, the thermal contribution 
dominates the invariant yield upto $p_T$ $\sim$ 4 GeV. The spectrum 
beyond $p_T >$ 4 GeV is dominated by pQCD calculations clearly 
distinguishing thermal window $1.5 < p_T(GeV)<4.0 $. 
The solid line in Fig.~\ref{fig5} displays the sum of thermal and 
pQCD contributions explaining the data within the ambit of considered 
parameters. 

\begin{figure}
\begin{center}
\includegraphics[scale=0.43]{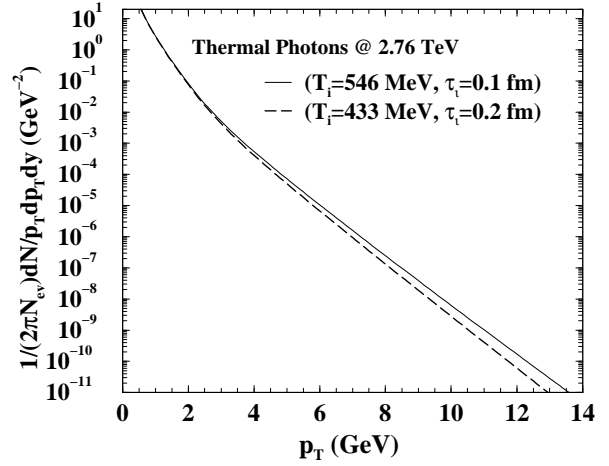}
\caption{Sensitivity in the production of thermal photons to $T_i$ for 
LHC energy. The long dashed line represents the results for $T_i$=433 MeV and 
$\tau_i$=0.2 fm. Solid for $T_i$=546 MeV and $\tau_i$=0.1 fm } 
\label{fig6}
\end{center}
\end{figure}

\begin{figure}
\begin{center}
\includegraphics[scale=0.43]{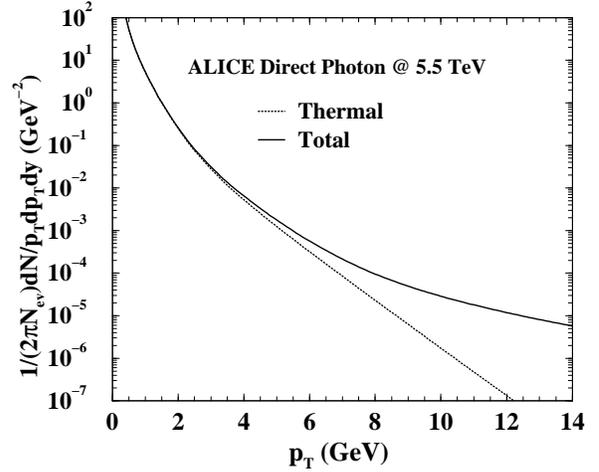}
\caption{Direct photon procutions at $\sqrt{s_{NN}}$=5.5 TeV LHC energy.} 
\label{fig7}
\end{center}
\end{figure}

It is important to mention the sensitivity of the results due to the 
initial parameters. The transition temperature $T_c$=163 MeV is restricted 
from the results of lattice QCD computation by HotQCD and Bielefield group
~\cite{hotqcd,bielefield}. However the computation as given in 
~\cite{latticetifr} predicts the value around $\sim$ 170 MeV. To differentiate 
the effect of $T_c$ we evaluate the photon spectra for LHC energy with 
$T_c$=170 MeV and found very little change in the $p_T$ spectra. At 
$p_T$=2 GeV, the results for $T_c$=170 MeV is  $\sim$ 1.2\% higher than 
$T_c$=163 MeV.
The sensitivity of $T_c$ is also found to be small in our earlier work
~\cite{JA07}. The sensitivity of results to the initial temperature  
(in GeV) and thermalisation time (in fm) ($T_i, \tau_i$) together is shown 
in Fig.~\ref{fig6}. Comparing the multiplicity we consider 
($T_i=433, \tau_i=0.2$) and reevaluate the photon spectra. The solid line is 
for ($T_i=546,\tau_i=0.1$) and the long-dashed line is for 
($T_i=0.433,\tau_i=0.2$). At $p_T$=2 GeV the result differs by 7.54\%. 
But the results are found to be sensitve to $T_f$. 
In Fig.~\ref{fig7} the prediction for direct photons at 
$\sqrt{s_{NN}}$=5.5 TeV is given with an initial QGP temperature 
$T_i$=843 MeV and $\tau_i$=0.08 fm. The charged particle multiplicity 
is taken to be 2500 for 0-5\% centrality. It shows the dominance of 
thermal photons up to $p_T$=4 GeV of the direct photon spectra.
\par
\begin{figure}
\begin{center}
\includegraphics[scale=0.43]{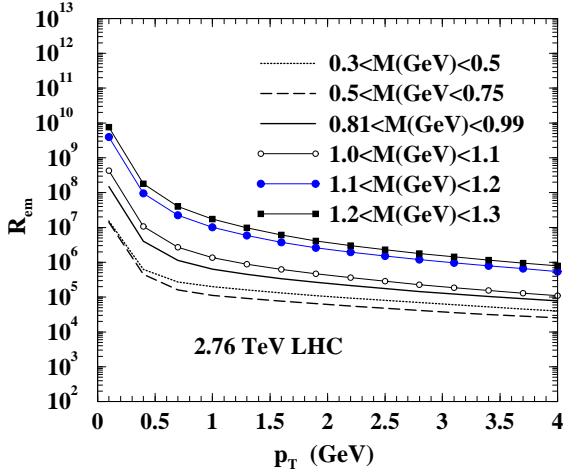}
\caption{Ratio of the invariant yield of thermal photons to dileptons  
at $\sqrt{s_{NN}}$=2.76 TeV LHC energy for different invariant mass windows.}
\label{fig8}
\end{center}
\end{figure}

It has been argued since long ~\cite{bsinha,ds}, that the ratio 
of the $p_T$ spectra of thermal photons to lepton pairs forms a plateau in 
the $p_T$ window $1.5 < p_T (GeV)< 3.5 $ for different mass bins of lepton 
pairs and 
it is less sensitive to the input parameters. In ref~\cite{JKN} the ratio 
has been revisited and the sensitivities to different input parameters 
have been studied extensively. Here we study the ratio for 
$\sqrt{s_{NN}}$ =2.76 TeV LHC energy for different mass window of the 
lepton pairs and plot the results in Fig.~\ref{fig8}. The thermal dilepton 
yield is calculated using the approach mentioned in ~\cite{ahns,sabyaepjc1} , 
that explains the dimuon data from In+In collisions measured by 
NA60 collaboration. A plateau is observed beyond $p_T$ $\sim$ 1.5 GeV 
in Fig.~\ref{fig8}.
\par
The plateau in the ratio appears because of the following reason.
The emission rate of photon and dilepton production show their momentum 
dependencies through their thermal phase space factor 
$f_{BE}(E,T)$ ($\sim exp(-E/T)$). Where $E=M_Tcoshy$, 
$M_T=\sqrt{p_T^2+M^2}$ \& $y=tanh^{-1}{p_z/E}$. $p_z$ and $y$ take their 
usual meanings. The nature of the spectra depends on $M$. When we consider 
the ratio of their momentum spectra at high $p_T$ 
($p_T>>M, \&  M_T=p_T$), 
for a static system, then the real photon 
($M^2=0$) and dilepton ($M^2\neq 0$) have similar momentum dependence. 
Thus a plateau is observed in ratio for the static system. However for a 
realistic case of an expanding system the thermal phase space factor goes 
as $f_{BE} \sim exp(-u_{\mu}p^{\mu})$, where $u_{\mu}$ and $p_{\mu}$ 
are the 4-velocity and -momentum. The spectra, then, depend on the 
radial flow $v_r$ along with $M$. It has been observed in~\cite{JKN} 
that for (i) $p_T >>M$, the radial flow of photons and dileptons are 
similar, and thus the plateau is achieved. (ii) If the large $M$ pairs 
originate from early time, when the flow is small, the ratio which 
includes the space time dynamics will also be close to a static case 
showing a plateau. (iii) But the plateau dissapears when ($p_T \sim M$) 
or radial flow is large. Here we present the ratio for different 
lepton pair mass windows for the range of $0.3 < M(GeV) <1.3$ at 
$\sqrt{s_{NN}}$=2.76 TeV.

It may be noted that the ratio of photons to dileptons obtained from pQCD 
calculation (non-thermal origin) doest not show a plateau, which is analysed 
extensively in~\cite{JKN}. 

In summary, we say, that the measured direct photon spectra from Au+Au 
collisions at 
$\sqrt{s_{NN}}$=200 GeV for 0-5\% centrality and from Pb+Pb collisions 
at $\sqrt{s_{NN}}$=2.76 TeV for 0-40\% centrality have been reproduced 
by ideal relativistic hydrodynamics with an EOS which represents crossover 
like transition. Initial QGP phase is assumed for both RHIC ($T_i$=404 MeV, 
$\tau_i$=0.2 fm) and LHC ($T_i$=546 MeV, $\tau_i$=0.1 fm). 
\par
It is observed 
somewhat remarkably that over a wide range of energies going through 
orders of magnitude, from (even SPS) RHIC to LHC, the thermal photons 
populate the same $p_T$ window $1.5 \le p_T(GeV) \le 3 $. 
The ratio $\gamma/\mu^+\mu^-$ for the invariant mass widow $0.3<M(GeV)<1.3$ 
and $p_T$ window $1.5\le p_T(GeV)\le 3$, turns to a plateau 
indicating the onset of QGP in the heavy ion collision. 
The formation of thermal window is manifested in the ratio 
within $1.0< p_T (GeV) <4.0$ at $\sqrt{s_{NN}}$=2.76 TeV LHC energy. 

{\bf Acknowledgment:} 
JKN thanks P. Mohanty, R. Sahoo, P. Tribedi, N. R. Sahoo, M. Mukherjee, 
S. Jena, S. Basu, P. Ghosh, T. Nayak and J. Alam for useful discussion. 
JKN also thanks Swagato Mukherjee of BNL for the discussion on recent 
lattice data. BS thanks J. Alam, S. Raha, L. Mclerran and DAE for Homi 
Bhabha Chair and the grant associated with it.

\end{document}